\documentclass{article}
\usepackage{smc}
\usepackage{times, amsmath,}
\usepackage{ifpdf}
\usepackage[english]{babel}
\usepackage{cite}
\usepackage[table]{xcolor}

\usepackage{color}
\usepackage{pifont}

\def\papertitle{StyleWaveGAN: Style-based synthesis of drum sounds with extensive
	controls using generative adversarial networks}
\def\firstauthor{Antoine Lavault}
\def\secondauthor{Axel Roebel}
\def\thirdauthor{Matthieu Voiry}

\newif\ifpdf
\ifx\pdfoutput\relax
\else
   \ifcase\pdfoutput
      \pdffalse
   \else
      \pdftrue
\fi

\ifpdf 
  \usepackage[pdftex,
    pdftitle={\papertitle},
    pdfauthor={\firstauthor, \secondauthor, \thirdauthor},
    bookmarksnumbered, 
    pdfstartview=XYZ 
   ]{hyperref}

  \usepackage[pdftex]{graphicx}
  \graphicspath{{./figures/}}
  \DeclareGraphicsExtensions{.pdf,.jpeg,.png}

  \usepackage[figure,table]{hypcap}

\else 
  \usepackage[dvips,
    bookmarksnumbered, 
    pdfstartview=XYZ 
  ]{hyperref}  

  \usepackage[dvips]{epsfig,graphicx}
  \graphicspath{{./figures/}}
  \DeclareGraphicsExtensions{.eps}

  \usepackage[figure,table]{hypcap}
\fi

\hypersetup{
    colorlinks,%
    citecolor=black,%
    filecolor=black,%
    linkcolor=black,%
    urlcolor=black
}

\title{\papertitle}

 \threeauthors
   {\firstauthor} {Apeira Technologies\\ STMS, Sorbonne Université\\%
     {\tt \href{mailto:a.lavault@apeira-technologies.fr}{a.lavault@apeira-technologies.fr}}}
   {\secondauthor} {STMS, IRCAM, CNRS\\Minstère de la Culture\\ %
     {\tt \href{mailto:roebel@ircam.fr}{roebel@ircam.fr}}}
   {\thirdauthor} {Apeira Technologies \\ %
     {\tt \href{mailto:m.voiry@apeira-technologies.fr}{m.voiry@apeira-technologies.fr}}}

\begin{document}
\capstartfalse
\maketitle
\capstarttrue
%


	\begin{abstract}
	
	In this paper we introduce StyleWaveGAN, a style-based drum sound generator
	that is a variation of StyleGAN, a state-of-the-art image generator 
	\cite{Karras2018,Karras2019}. By conditioning StyleWaveGAN on both the type of drum and several audio descriptors, we are able to synthesize waveforms
	faster than real-time on a GPU directly in CD quality up to a duration of 1.5s
	while retaining a considerable amount of control over the generation. We also introduce
	an alternative to the progressive growing of GANs and experimented on the effect
	of dataset balancing for generative tasks. The experiments are carried out on an
	augmented subset of a publicly available dataset comprised of different drums
	and cymbals. We evaluate against two recent drum generators,
	WaveGAN \cite{Donahue2018} and NeuroDrum \cite{Ramires2019},
	demonstrating significantly improved generation quality (measured with the
	Frechet Audio Distance) and interesting results with perceptual features.

\end{abstract}

\section{Introduction}\label{sec:introduction}

Drum  machines are musical devices creating percussion sounds using analogue or digital signal processing \cite{Reid}\cite{Reida}. The characteristic sound of this synthesis process contributed to their use in the '80s and their appreciation nowadays. However, these drum machines did not provide an extensive set of controls over the generation.

Following the success of deep learning, several generative processes for
percussive sounds have been proposed in the recent years, and two approaches
retained our attention. \cite{Drysdale2020} used a GAN for waveform
generation with a conditioning on the type of drum, generating 0.3s at
44100kHz. There is also \cite{Nistal2020a}, where
a GAN was trained to generate STFT of drum sounds while controlling the
generator with audio descriptors, allowing them to generate 1s at 16kHz.
Both of them used the progressive growing of GANs \cite{Karras2017}.

In  this  paper,  we  build upon  the  same  idea  of  conditional  synthesis 
using discrete and continuous  controls, with time-domain generation like
\cite{Drysdale2020} and control by means of perceptual features derived from the
AudioCommons project like \cite{Ramires2019, Nistal2020a} with a style-based approach
(SGAN)\cite{Karras2018,Karras2019}. The charactheristics of these networks are summarized in table \ref{tab:charc_net}. We expand on the idea of control with perceptual features by means
of replacing the trained auxiliary network used in \cite{Odena2016, Nistal2020a}
with a  differentiable implementation of the feature estimators, increasing the robustness of the feature evaluation. We conduct our experiments on an augmented
version of the ENST-Drums \cite{Gillet2006} dataset, containing kick, snare,
toms and hi-hats and comprising about 120k samples amounting to 100 hours of
recordings. To evaluate the quality of the model on this dataset, we are using
the Fréchet Audio Distance (FAD) \cite{Kilgour2019}, in an attempt to obtain a
reference-free automatic evaluation of the generated samples. Finally, we explore the ability of the network to use the information from the perceptual features.

All in all, our goal is to create an algorithm for drum sound synthesis suitable for professional music production. In other words, we expect good output quality, real-time generation and relevant controls. The Fréchet Audio Distance (FAD) is used for the quality evaluation, real-time ability is measured through plain generation and the quality of the controls uses the descriptor consistency metric from \cite{Ramires2019}.

\begin{table}[htb]
	\begin{center}
		{%
			\begin{tabular}{|c|c|c|}
				\hline
				Reference & Sample Rate & Duration \\
				\hline
				WaveGAN \cite{Donahue2018}& 16kHz  & 1.1s\\
				NeuroDrum \cite{Ramires2019}& 16kHz  & 1s \\
				DrumGAN \cite{Nistal2020a}& 16kHz & 1.1s \\
				Drysdale et al.\cite{Drysdale2020}& 44.1kHz  & 0.4s \\
				\hline
				\textbf{Ours} & \textbf{44.1kHz} & \textbf{1.5s}\\
				\hline
			\end{tabular}
		}
	\end{center}
	\label{tab:charc_net}
	\caption{Comparison of state of the art neural drum synthesizers}

\end{table}

\section{Model}
\subsection{Audio-Commons Timbre Models}
The Audio Commons project implements a collection of perceptual models that describe high-level timbral characteristics of a sound \cite{Pearce2016}. These features are specially crafted from the study of popular timbre designations given to a collection of sounds from the Freesound dataset. The perceptual models are built by combining existing low-level features found in the literature \cite{Peeters2004}, which correlate with the chosen timbral designation.

Contrary to \cite{Nistal2020a}, we reimplemented those features in order to make them fit directly into the
training as differentiable functions. Our motivation behind this comes from the use of an auxiliary network for	conditioning in \cite{Nistal2020a}. Constructing a differentiable proxy for
these timbral features by training a neural network does not guarantee the correct evaluation of the features to the same degree than implementing the features following the reference implementation. Moreover the  direct implementation allows a correct evaluation of signals that have descriptor values outside of  the range of values that were available for training the proxy. Our implementation of these descriptors as well as the supplementary material can be found at \url{https://alavault.github.io/stylewavegan/}

\subsection{Generative Adversarial Networks and StyleGAN}

Generative Adversarial Networks (GAN) are a family of training  procedures
in  which  a generative model (the generator) competes against a
discriminative adversary (the discriminator) that learns to
distinguish whether a sample is real or fake \cite{Goodfellow2014}.

Instead of using a vanilla GAN, we are using an evolution called StyleGAN
\cite{Karras2018, Karras2019}. StyleGAN attempts to mitigate the entangled
representation when using noise as latent and input of the
generator. The key idea here is to use a \textsl{style encoding}, a vector which is obtained through a mapping network and is then used to	control (through an affine transform) every layer of a synthesis network.

\subsection{Proposed architecture}

\begin{figure}[htb]
	\centering
	\includegraphics[width=0.65\columnwidth]{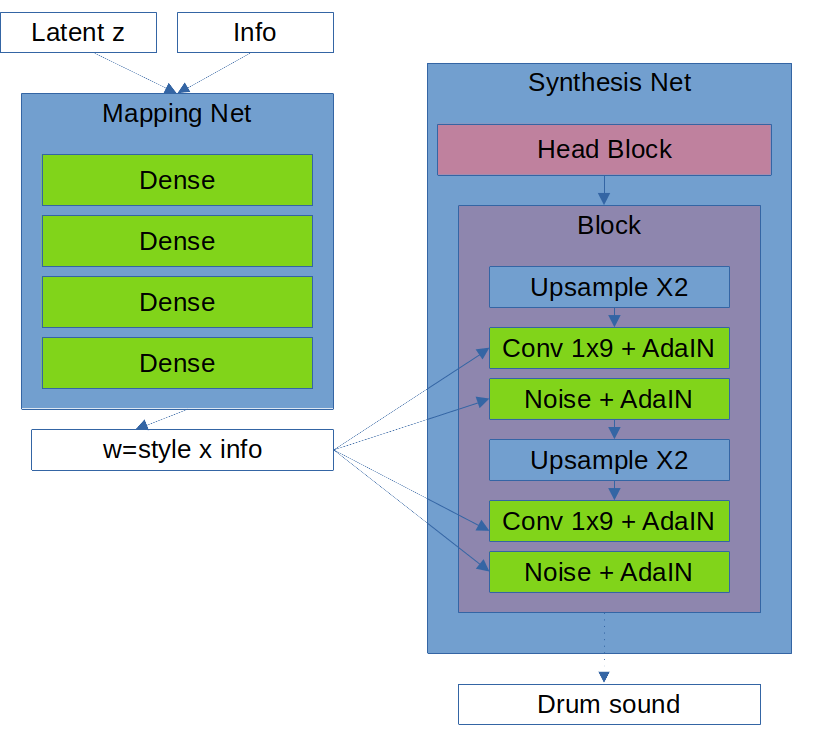}
	\caption{StyleWaveGAN}
	\label{fig:stylewavegan}
\end{figure}

Since StyleGAN was originally used for high-quality image generation, we have
to modify it for direct waveform generation. In particular, we
transform 2D convolution ($3 \times 3$) into 1D
causal convolutions ($1 \times 9$) \cite{Oord2016}, the upsampling is
done with an averaging filter before each convolution block in the synthesis
network, the mapping networks has 4 layers instead of 8 and the loss function is
WGAN-LP \cite{Petzka2017} (see figure \ref{fig:stylewavegan}).

We use the same number of filters, with respect to the depth, as StyleGAN2\cite{Karras2019}. Just
like StyleGAN2 , the synthesis network uses input/output skips
and the discriminator is a residual network.

In this work we follow \cite{Ramires2019,Drysdale2020} using a temporal signal
representation. Informal perceptual evaluations performed in the initial phase
of this study supported our idea that the temporal representation produces
better audio quality than spectral representation : we suppose it is because of the high amount of noise and the importance of the transient in the drum sounds.  


\subsection{Noise addition layers and output envelopes}

We modified the noise addition layers of StyleGAN to make them style-dependant. We also add noise shaping (with a linear fade out) to avoid noisy tails. Having controlled noise addition is useful since some classes need more noise than other to get a good quality synthesis.

This can be summarized in the following equation :

\begin{equation}
y = x + w\cdot n + b
\end{equation}
where $y$ is the output of the layer, $x$ is the signal input of the layer, $w$ is the transformed style vector, $n$ the shaped noise (the same on every channel) and finally $b$ a bias term.

One of the drawback of having noise addition layers is the lack of control of the decay of said noise. Because of this, the generated sounds have an audible noisy tail which makes them easily identifiable by a human listener. To avoid this pitfall, we added envelopes after the output of the network.

These envelopes where generated using the training dataset, one per type of drum. For each sample of one given type, the final envelope is the filtered mean of the analytical part of the Hilbert transform of these normalized samples. A small fade out is applied to avoid audible clicks at the end of the generated sounds. The Hilbert transform is calculted using the Discrete Fourier Transform on the first 1.5s (65636 samples at 44.1kHz) of each normalized sound of the dataset.

\begin{figure}[htb]
	\centering
	
	\includegraphics[width=1\columnwidth]{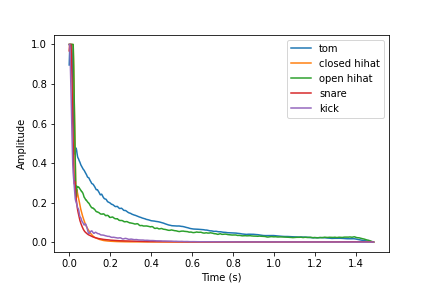}
	\caption{Generated envelopes from the training dataset}
	\label{fig:envelope}
\end{figure}

The final audio is obtained by multiplying the output of the synthesis network and the matching envelope element-wise. This ensure a quasi-constant energy representation inside the synthesis network. We hypothesize this helps by reducing the dynamic range to generate by the non-linearities inside the network.

The output time signal of the network $y_n$ is obtained from the output $x_n$ of the network by means of multiplying the envelope signal $e_{n,c}$ for drum class $c$  by means of 
\begin{equation}
y_n = x_n e_{n,c}
\end{equation}

\subsection{Controlling the network}

The labels and audio descriptors are fed into an embedding layer
which is then concatenated to the latent $z$ (c.f figure \ref{fig:stylewavegan})
and fed to the mapping network. These labels and descriptors are concatenated
after the mapping network too.

In our experiments, we are using 5 labels. These labels are added to the network with a one-hot vector. The
descriptors, if used, are concatenated after the labels. We expect to have	a better disentanglement between the class label or the descriptors during the style encoding by using this method. We use the L1 loss to measure the deviation between the target descriptors and the generated values.

\subsection{AutoFade}

Progressive growing of GANs has been proposed in \cite{Karras2017} and used in \cite{Drysdale2020, Nistal2020a}. In our experiments, we developed and evaluated a variant of progressive growing, that we denote AutoFade. It is a ResNet architecture with a
convolution path and a bypass where a learned parameter is used to fade more or less of one path. Rather
than fixing a value like ResNet, we let the network choose the best value as
part of the training process, without the need of training it block by block. 
If $x$ and $y$ represents the two different branches, we have:

\begin{equation}
\sin(\alpha) x + \cos(\alpha)y
\label{eq:autofade}
\end{equation}

$\alpha$ is independant of $x$ or $y$. It makes this structure an
intermediate between ResNet and Highway Networks. By using trigonometric function in equation (\ref{eq:autofade}), we guarantee the conservation of the standard deviation, if both inputs have equal variance. Similar to \cite{Karras2019} we did not find any benefit using progressive growing or AutoFade in the generator. On the other hand using progressive growing in the discriminator did improve the results. The Autofade feature will therefore be evaluated in the following sections, only as part of the discriminator.

\section{Experimental setup}

\subsection{Dataset }

We are using a subset of ENST-Drums \cite{Gillet2006}, comprised of 350 samples
of close miking of kicks, snares, toms and hi-hat. 	Since 350 elements is too low for a data-driven approach, we used an augmentation method similar to \cite{Jacques2019a}. We used
\texttt{SuperVP} \footnote{SuperVP is available free of charge  in form of a
	Max/MSP object at \url{https://forum.ircam.fr/projects/detail/supervp-for-max/} } to process the original dataset. 
The modifications applied to the sounds consist of a gain applied to transient/attack components \cite{Roebel2003}, noise components as well as independent transposition of the signal source and the spectral envelope. 

The set of parameters is shown in table \ref{tab:augmentation}. The limits have been obtained by means of subjective evaluation of the modified sounds aiming to avoid transformations that can be perceived as unnatural by a human listener. Examples are available in the supplementary material. 

As a supplementary metric, the Fréchet Audio Distance between the original dataset and the augmented one is 0.62.


\begin{table}[htb]
	\begin{center}
		\begin{tabular}{|l|l|}
			\hline
			Process & Parameters \\
			\hline
			Remix attack  & 0.1, 0.3, 0.6, 1.5, 2, 3 \\
			Remix noise  & 0.6, 1.5, 2, 3 \\
			Transposition  & 0, $\pm$100, $\pm$200 \\
			Spectral envelope transposition  & 0, $\pm$200 \\									
			\hline
		\end{tabular}
	\end{center}
	\caption{Augmentation operations and parameters}
	\label{tab:augmentation}
\end{table}


\subsection{Training procedure}

The training procedure is the same as StyleGAN 2 \cite{Karras2019}, except that
we trained the network on 2M samples. With a batch size of 10, it totals
to 200k iterations.

\subsection{Imbalanced dataset}
\label{sec:imbalanced}
Balancing datasets is common in classification tasks but to our knowledge, has never been done for generation tasks. As shown in table \ref{tab:population}, our augmented dataset is quite unbalanced, so to obtain a balanced dataset, we use a sampler which takes elements from sub-datasets (one per label) at random according to a uniform distribution. We call it "equal-proportion sampling".

\begin{table}[htb]
	\begin{center}
		\begin{tabular}{|l|l|}
			\hline
			Element & Proportion \\
			\hline
			Kick  & 3\% \\
			Snare  & 18\% \\
			Toms  & 45\% \\
			Closed hi-hat  & 10\% \\
			Open hi-hat  & 22\% \\
			\hline
		\end{tabular}
	\end{center}
	\caption{Dataset population}
	\label{tab:population}
\end{table}


\subsection{Baseline}
\label{sec:experimental_results}
The most appropriate candidate to be used as our baseline is DrumGAN \cite{Nistal2020a} and \cite{Drysdale2020}.
Unfortunately, these are not reproducible because of missing source code or/and
missing or unknown meta parameters. Therefore, we will compare to
\cite{Ramires2019} using the distributed code and a reimplementation of
\cite{Donahue2018}, both trained on our augmented dataset.

Because NeuroDrum \cite{Ramires2019} works with 16kHz samplerate we adapted our model to use this sample rate for this comparison. We also compared with WaveGAN \cite{Donahue2018} using our dataset with 44.1kHz. Here we configured both networks to
generate 0.3s (@44.1kHz).


\subsection{Evaluation}

We chose to use the Fréchet Audio Distance (FAD) \cite{Kilgour2019}, a
reference-free evaluation metric for audio generation algorithms using a VGGish
model trained on AudioSet. We compare the embedding of the augmented database to
the embedding obtained from 64k samples generated by the evaluated network. In terms
of computational cost, we achieve a generation rate of $52\text{drum sounds}/s$ on
one 1080GTX with the network in full resolution (1.5s@44.1kHz + descriptors).


\begin{table}[htb]
	\begin{center}
		\begin{tabular}{|l|l|}
			\hline
			Network & FAD \\
			\hline
			Baseline \cite{Ramires2019} & 25.35\\
			\textbf{StyleWaveGAN@16kHz}  & \textbf{11.48} \\
			\hline
		\end{tabular}
	\end{center}
	\caption{FAD comparison to NeuroDrum \cite{Ramires2019} (lower is better)}
	\label{tab:fad_neurodrum}
\end{table}

\begin{table}[htb!]
	\begin{center}
		\begin{tabular}{|l|l|}
			\hline
			Network & FAD \\
			\hline
			Baseline@44.1kHz \cite{Donahue2018}  & 13.08 \\
			\textbf{StyleWaveGAN@44.1kHz} (SWG)  & \textbf{7.75} \\
			\hline
			\textbf{SWG + AutoFade} (AF)  & \textbf{6.84}\\
			SWG + Balanced dataset (B)  & 7.89  \\
			SWG + AF + B  & 7.92 \\
			\hline
		\end{tabular}
	\end{center}
	\caption{FAD on networks without labels (lower is better)}
	\label{tab:fad_1}
\end{table}

\begin{table}[htb!]
	\begin{center}
		\begin{tabular}{|l|l|}
			\hline
			Network & FAD \\
			\hline
			SWG + labels & 6.85 \\
			SWG + labels + AF  & 6.72\\
			SWG + labels + AF + Balanced data (B)  & 6.65\\ 

			\textbf{SWG + labels + AF + B + Envelope}  & \textbf{3.62}\\ 
			
			\hline
		\end{tabular}
	\end{center}
	\caption{FAD on label-conditioned networks (lower is better)}
	\label{tab:fad_label}
\end{table}

\begin{table}[htb!]
	\begin{center}
		\begin{tabular}{|m{0.2\columnwidth}|m{0.2\columnwidth}|m{0.2\columnwidth}|m{0.2\columnwidth}|}
			\hline
			Class & SWG &SWG + AF + B & SWG + AF + B + Env  \\
			\hline
			Kick       & {8.79}  & 11.71    & \textbf{3.58}\\
			Snare      & 7.87    & {7.53}   & \textbf{4.29}\\
			Tom        & 8.17    & 8.09     &  \textbf{6.27}\\
			Closed HH  & 10.12   & 6.97     & \textbf{4.23}\\
			Open HH    & 8.26    &  8.91    & \textbf{4.12} \\
			\hline
		\end{tabular}
	\end{center}
	\caption{Intra-class FAD for label-conditioned StyleWaveGAN}
	\label{tab:fad_2}
\end{table}

\section{Experimental results}

This section describes the results obtained with StyleWaveGAN on three main configurations. The First unconditioned, the second with conditioning on the labels and finally a third with labels and descriptors.

\subsection{Impact of our contributions}

The first result for unconditioned synthesis we have is that we improved on our baseline in terms of FAD (tables \ref{tab:fad_neurodrum} and \ref{tab:fad_1}). We can also see from table \ref{tab:fad_1} that using AutoFade in the discriminator helped at getting a better generation in this context.

The results with dataset balancing are mitigated.  Without the label conditionning, using it didn't bring any
decrease in the FAD : since it makes the training and evaluation dataset different (in proportions), the learned distribution differs, impacting negatively the FAD. This can be seen in table \ref{tab:fad_1}. However, it improved the supervised generation, as seen on table \ref{tab:fad_label}.

The impact on the intra-class FAD of AutoFade and dataset balancing is shown in table \ref{tab:fad_2}. It lowers the FAD generally except for the kick and open hihat. Output envelopes have a very strong impact on the FAD for all drum classes. They reduce the FAD by nearly two for all drum classes besides for the tom.

\subsection{Control with audio descriptors}

We will investigate further on the control of perceptual features. We trained a
network on the same dataset, but we made it generate longer audio : 65536
samples, equivalent to 1.48s. Examples are available in the supplementary material.

\subsubsection{Brightness}

\begin{figure}[htb]
	\centering
	
	\includegraphics[width=1\columnwidth]{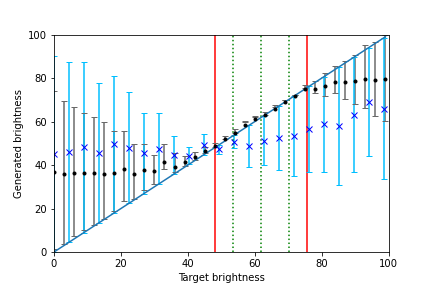}
	\caption{Target brightness vs. Generated brightness (single descriptor)}
	\label{fig:brightness}
\end{figure}

We only focus on one class (snare) and one	descriptor (brightness) as a first presentation of the idea. Figure \ref{fig:brightness} shows the relation between	target and synthesized brightness for NeuroDrum and StyleWaveGAN. Results are  shown in form of  mean values and standard  deviation in black dots
(StyleWaveGAN) and blue crosses (NeuroDrum). The solid red vertical lines show
the limiting values in the  training dataset. Finally, the reference target
values used for the  ordering comparison according to
\cite{Ramires2019,Nistal2020a} and discussed below are marked with dotted green
lines.
This figure demonstrates  clearly  that  while  the  mean  value  of  the
perceptual brightness  of a sound  produced by  NeuroDrum is
increasing with  the target  brightness, it still  remains far  off the target 
brightness most  of  the time. In  contrast, the  synthesized
brightness of StyleWaveGAN  is very close to the target  value for all values
that are present in the  training set and even remains somewhat close to the target
outside the brightness limits of the training data. 

To compare to \cite{Ramires2019,Nistal2020a}, we are using the ordering criterion used in \cite{Ramires2019,Nistal2020a}. It compares pairs of sounds generated with a pair of target values (situated at levels 0.2, 0.5 and 0.8 on a min/max normalized scale), and	evaluates whether the ordering of the targets is  preserved in the generated features. Like \cite{Ramires2019}, E1 uses extreme points, E2 uses the mid and low values and E3 uses the mid and high values. The very small error in the synthesized feature values generated with StyleWaveGAN results in a consistent ordering for all three criteria. Table \ref{tab:nistal_repro} reproduces the results for brightness control from table 3 in \cite{Nistal2020a} comparing NeuroDrum and DrumGAN, trained on a different
dataset under the column "D1". The results under the columns "D2" are for our network, trained on our augmented dataset. We matched and improved the results from NeuroDrum and DrumGAN in this configuration. 

All these  results support our  hypothesis that replacing  a trained
feature  estimator  as  in \cite{Nistal2020a,Odena2016} by  means  of  a 
direct	implementation of the feature  estimator allows for a significantly improved control
consistency of the final network.

\begin{figure}[htb]
	\centering
	
	\includegraphics[width=1\columnwidth]{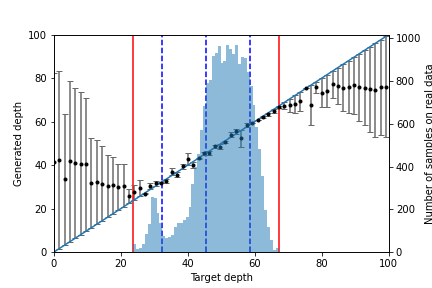}
	\caption{Target depth vs. Generated depth (single descriptor)}
	\label{fig:depth}
\end{figure}
\begin{figure}[htb]
	\centering
	
	\includegraphics[width=1\columnwidth]{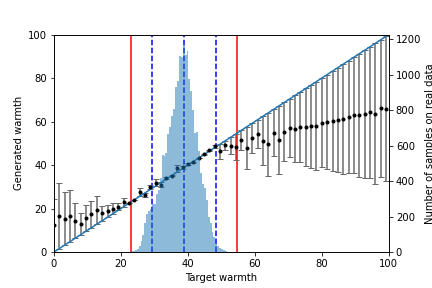}
	\caption{Target warmth vs. Generated warmth (single descriptor)}
	\label{fig:warmth}
\end{figure}

\begin{table}[htb]
	\begin{center}

		\begin{tabular}{|p{2cm}|c|c|c|c|c|c|c|c|}
			\hline
			Features & \multicolumn{2}{c|}{E1} &
			\multicolumn{2}{c|}{E2}&\multicolumn{2}{c|}{E3}\\
			\hline
			Dataset& D1 & D2 &  D1 & D2 & D1 & D2 \\
			\hline
			DrumGAN & 0.74 & - & 0.71 & - & 0.7&- \\
			\hline
			NeuroDrum & 0.99 & 0.91 & 0.99 & 0.80 &0.99 & 0.68 \\ 
			\hline
			SWG & - & 1.00 & - & 0.94 &- & 0.98 \\
			\hline
		\end{tabular}
		
	\end{center}
	\caption{Ordering accuracy for the feature coherence tests for brightness on  samples  generated	with  the  baseline  NeuroDrum \cite{Ramires2019}  and DrumGAN (from
		\cite{Nistal2020a}), higher is better}
	\label{tab:nistal_repro}
\end{table}

\subsubsection{Other descriptors}

We will discuss here the results for some other descriptors we deem of interest for our task : depth and warmth. Results are shown in table \ref{tab:descriptors} as well as figures \ref{fig:depth} and \ref{fig:warmth}. On these figures, an histogram of the dataset values is overlayed in light blue.

Figure \ref{fig:depth} shows the results for the depth descriptor. We have a slight worse performance than the brightness descriptor due to some outliers. The same extrapolation property is found here, but slightly less smooth. We can conclude that the depth descriptor is harder to learn for the network. The difference for low depth ($<30$, marked by the first blue dashed line) can be explained by the low number of samples to train the network with at this level.

Figure \ref{fig:warmth} shows the results for the warmth descriptor. The performance is on par with the brightness descriptor except for the region above 80\% of the min/max value. This can be explained by a lack of training data is this region as shown on the overlaid histogram.

\begin{table}[h!]
	\begin{center}
		\begin{tabular}{|l|l|l|l|}
			\hline
			Features & E1 & E2 & E3  \\
			\hline
			Depth &0.99 & 0.99 &0.71
			\\
			Warmth &1.00 & 0.86 & 0.90
			\\
			\hline
		\end{tabular}
	\end{center}
	\caption{Ordering accuracy for other features of interest using StyleWaveGAN (higher is better)}
	\label{tab:descriptors}
\end{table}

\subsubsection{Multi dimensional descriptor controls}

Using three individual networks for controlling the indvidual descriptor is not that interesting for a real world application.  
In the next step we therefore investigate controlling the network with a 3 dimensional vector of warmth, depth and brighness descriptors. 

When  using descriptors simultaneously as part of the control, we can expect conflicts between them as well as dependence to the training data. Since the network is trained on data, it will learn to reproduce similar features as the real data which also means only a part of the combination possibles.

To evaluate the quality of control, we use the same label but change the evaluation method slightly. While we use the same criterion, we generate samples in a way that can create sounds outside of the training dataset. More precisely, we take a set of real features from a batch of the training data and then modify the descriptor to be evaluated to 20, 50 or 80 percent of the min/max value with respect to said descriptor. Results obtained using this method are shown in table \ref{tab:descriptors_together}.

\begin{table}[h!]
	\begin{center}
		\begin{tabular}{|l|l|l|l|}
			\hline
			Features & E1 & E2 & E3  \\
			\hline
Brightness & 1.0 & 1.0& 1.0\\
Depth& 1.0 & 1.0& 0.99 \\
Warmth & 0.98 & 0.59 & 0.97 \\
			\hline
		\end{tabular}
	\end{center}
	\caption{Ordering accuracy for multiple descriptors using StyleWaveGAN (higher is better)}
	\label{tab:descriptors_together}
\end{table}



As shown in table \ref{tab:descriptors_together}, training the descriptors with the proposed differentiable  error function produces a network following controls with a precision such that the ordering criterion proposed in \cite{Ramires2019} and used in \cite{Nistal2020a} is no longer sufficient to evaluate the control precision. In the following we therefore propose a refined evaluation criterion that allows evaluating control precision with more details, taking into account not only ordering but also errors.

In order to achieve this, we will be using the Mean Absolute Error (MAE) between the target values and the output values on three regions based around quantiles of the dataset values :

\begin{itemize}
	\item F1 : MAE evaluated using only the target descriptor values within the 20th and 50th quantiles  
	\item F2 : MAE evaluated using only the target descriptor values within the 50th and 80th quantiles 
	\item F3 : MAE evaluated using only the target descriptor values within the 20th and 80th quantiles 
\end{itemize}

First, the interest of working with quantiles rather than percentage of the min/max values is that we expect to cover the same amount of values of the dataset each time while avoiding extreme values. The results are shown in table \ref{tab:descriptors_mtge}. The values given in said table are not percentage or relative to the descriptor values : they are absolute errors.  We can also note that these numbers have the same unit as the descriptors.

In table \ref{tab:descriptors_mtge}, the lines labelled \textsl{single} show the results using networks with only one descriptor and the lines labelled \textsl{combined} show the results when the descriptor of interest is set but the others are taken from a real sound from the training dataset. Finally, the lines labelled \textsl{combined, dataset} show the results when all the descriptors values are taken from the training dataset.

\begin{table}[h!]
	\begin{center}
		\begin{tabular}{|l|l|l|l|}
			\hline
			Features & F1 & F2 & F3 \\
			\hline
			NeuroDrum (brightness) &7.22 & 10.40 & 8.81 \\		
			Brightness (single) & 0.83 &1.06 &0.98 \\
				
			Depth (single) & 1.06& 1.15& 1.10\\	
			Warmth (single) & 1.15 &1.01 & 1.08 \\	
			\hline
			Brightness (combined) & 0.97 &1.36 &1.17 \\
			Depth (combined) & 1.33 &1.50 &1.41\\
			Warmth (combined) & 1.29 & 3.31& 2.33\\
			\hline
				Brightness (dataset, combined) & 0.75 & 0.95 & 0.85 \\
	Depth (dataset, combined) & 0.99 & 1.03 & 1.0\\
	Warmth (dataset, combined) &  1.42 & 1.37& 1.39\\
	\hline
		\end{tabular}
	\end{center}
	\caption{Mean absolute error for several configurations (lower is better)}
	\label{tab:descriptors_mtge}
\end{table}

Since we aim consider a perfect output follows perfectly the control input, we expect to see a good linear fit on the output. To evaluate this, we will calculate a linear least-square regression on the domain bound by the 20th and 80th quantiles, and use its determination coefficient $R^2$ as a metric of good linearity. In this case, $R^2$ is equal to :

\begin{equation}
R^2 = 1-{\frac{\sum_{i=1}^{n} (y_{i}-{\hat {y_{i}}})^2} {\sum _{i=1}^{n} (y_{i}-{\bar{y}})^2}}
\end{equation}

where $n$ is the number of samples, $y_i$ is the output value of the $i$-th measure, $\hat{y_{i}}$ the corresponding predicted value and $\bar{y}$ the average of the measured values. The results are compiled in table \ref{tab:descriptors_r2}. We can also note that we can use the slope from the least-square regression and use it as an ordering criterion.

\begin{table}[h!]
	\begin{center}
		\begin{tabular}{|l|l|}
			\hline
			Features & $R^2$  \\
			\hline
			NeuroDrum (brightness) &0.03  \\		
			Brightness (single) & 0.75  \\
			Depth (single) &  0.70 \\	
			Warmth (single) & 0.76 \\		
			\hline
			Brightness (combined) & 0.47 \\
			Depth (combined) & 0.67 \\
			Warmth (combined) & 0.08\\
			\hline
			Brightness (dataset, combined) & 0.72\\
			Depth (dataset, combined) & 0.62 \\
			Warmth (dataset, combined) &  0.45\\
			\hline
		\end{tabular}
	\end{center}
	\caption{Determination coefficient for several configurations (higher is better)}
	\label{tab:descriptors_r2}
\end{table}

\begin{figure}[htb!]
	\centering
	
	\includegraphics[width=1\columnwidth]{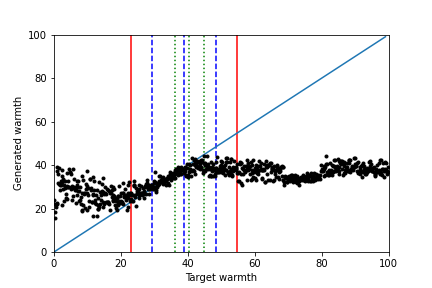}
	\caption{Target warmth vs. Generated warmth with combined descriptors}
	\label{fig:warmth3}
\end{figure}

Apart from a better fit than NeuroDrum, we can see that the $R^2$ coefficient is generally quite satisfying except for the warmth when used with values outside of the dataset. This illustrated on figure \ref{fig:warmth3}, where there is a bend in the output value.

This bend is due to the dataset value distribution, where for high warmth values, the set of values for the other descriptors gets small (a variation of less than 5 points around 50 for brightness and 66 for depth, these values being already quite rare in the dataset). So, when the control inputs gets brightness and depth values that are from the rest of the dataset, the warmth value has to be extrapolated by the network since such combination was not seen during training.

However, this behaviour is not shown when evaluating on control values from the dataset ($0.08 \leftrightarrow 0.45$). For the other descriptors, the linearity remains satisfaying whatever the evaluation method used.

\begin{figure}[htb!]
	\centering
	
	\includegraphics[width=1\columnwidth]{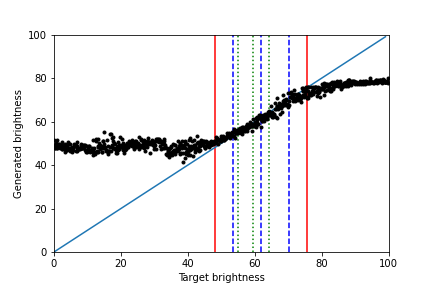}
	\caption{Target brightness vs. Generated brightness with combined descriptors}
	\label{fig:brightness3}
\end{figure}

\begin{figure}[htb!]
	\centering
	
	\includegraphics[width=1\columnwidth]{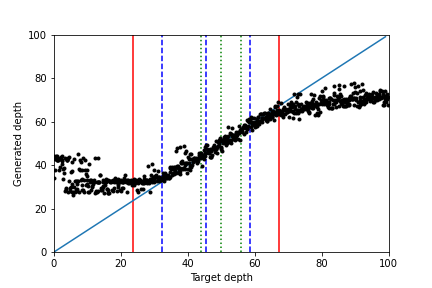}
	\caption{Target depth vs. Generated depth with combined descriptors}
	\label{fig:depth3}
\end{figure}

These considerations can be seen on the figures \ref{fig:brightness3} through \ref{fig:warmth3}. When iterating on the whole scale (i.e 0 to 100) while setting the other descriptors with values from the training dataset, the output control stays mostly consistent and linear and even allow to generate samples outside the minimum and maximum values of the dataset.


To conclude, we have justified our method works great almost everywhere in the min/max values of the training dataset and can extrapolate further than the min/max values as well as between unseen combination of descriptors.

\section{Conclusion and future work}

In  this paper, we presented  a new  method for  drum synthesis using
StyleWaveGAN, an adaptation of a state of the art image generator. The
proposed method has explicit controls  on drum type and  additional continuous controls for  selected perceptive audio features.

We  have shown  the  proposed style-based synthesis achieves a significantly reduced FAD compared to recent DNN based drum  synthesis  methods \cite{Donahue2018,Ramires2019}.  We  have  proposed a new  means for training  the  feature control  by  using  a  differentiable implementation of the AudioCommons   features  for
calculating the  feature loss and  have demonstrated that  this method
significantly improves  the feature  coherence between  target and measured features in the synthesized sounds when compared to \cite{Ramires2019}, and argue that the same improvement would hold  compared to \cite{Nistal2020a}. We also introduce a way to measure the fidelity of the control with respect to the input.
To the best  of our knowledge the proposed DNN  is the first achieving
drum synthesis with 44.1kHz sample rate (for sounds with a duration of
1.5s) with an  inference speed  more than  50
times  faster than  real time on a consumer GPU.

In terms of future work we will continue to work on the sound quality and
additional controls, notably regarding velocity.

%

%


\bibliography{Biblio_CIFRE.bib}

\end{document}